\documentclass[a4paper]{VisionStyle}
\usepackage{epsfig}

\begin{document}

\title{The nature of ultraluminous X-ray sources in nearby galaxies}

\author{T.P.\,Roberts\inst{1} \and M.R.\,Goad\inst{1}$^,$\inst{2} \and
M.J.\,Ward\inst{1} \and R.S.\,Warwick\inst{1} \and P.\,Lira\inst{1}} 

\institute{
Department of Physics \& Astronomy, University of Leicester,
University Road, Leicester, LE1 7RH, UNITED KINGDOM 
\and
Department of Physics \& Astronomy, University of Southampton,
Highfield, Southampton, Hants., SO17 1BJ, UK
}

\maketitle 

\begin{abstract}

The advanced capabilities of the {\it Chandra\/} and {\it
XMM-Newton\/} observatories mean that, for the first time, the
detailed study of the brightest point-like X-ray sources in nearby
galaxies outside of the local group is a realistic aim.  Here, we
present the results of a {\it Chandra\/} ACIS-S study of two of the
nearest and brightest sources in the rare ultraluminous (L$_{\rm X} >
10^{39} \rm ~erg~s^{-1}$) X-ray source class, \object{NGC 5204 X-1}
and \object{NGC 4559 X-1}.  When considered with new optical integral
field spectroscopy data this provides powerful diagnostics as to the
nature of these sources, in particular suggesting that NGC 5204 X-1 is
a high-mass X-ray binary, and showing new evidence linking it to the
Galactic microquasar phenomenon.  We also find that both ULX appear to
be located in cavities in emission-line gas nebulae that surround the
sources.  In addition, we present the results of a {\it Chandra\/}
observation of the interacting galaxies \object{NGC 4485}/\object{NGC
4490}, a pair of late-type spiral galaxies that, remarkably, contain a
total of six ULX.  We identify one as a supernovae, and the remainder
as probable black hole X-ray binaries.  All six are located in star
formation regions, underlining the emerging link between ULX and
active star formation activity.

\end{abstract}

\section{Motivation}

{\it EINSTEIN\/} imaging observations were the first to reveal that
the X-ray emission of a subset of spiral galaxies is dominated by one
or more very luminous (L$_{\rm X} \sim 10^{39-40}$ erg s$^{-1}$)
discrete X-ray sources located outside the nucleus of the galaxy
(\cite{troberts-E3:fab89} and references therein).  {\it ROSAT\/} and
{\it ASCA\/} observations have since confirmed the presence of many of
these ``ultraluminous X-ray sources'' (ULX)\footnote{A note on
taxonomy; we refer to the class of L$_{\rm X} > 10^{39}$ erg s$^{-1}$
extra-nuclear X-ray sources as ``ultraluminous X-ray sources''
throughout this proposal.  This class of object is alternatively
referred to as ``super-luminous X-ray sources'' (SLS),
``super-Eddington X-ray sources'' (SES) or ``intermediate-luminosity
X-ray objects'' (IXOs) in the literature.}  in nearby galaxies.  A
first indication of the statistical incidence of ULX in the nearby
universe came from the {\it ROSAT\/} HRI survey of bright, nearby
galaxies (\cite{troberts-E3:rw2000}) where we catalogued 28 sources
with L$_{\rm X} > 10^{39} \rm ~erg~s^{-1}$ (0.1 - 2.4 keV) in the
outer regions of a sample of 83 galaxies, with only one in five of the
galaxies surveyed hosting one or more ULX.  This clearly underlines
the rarity of this source class.

Given the high luminosities of this class of objects, which appear
intermediate between classic X-ray binaries (L$_{\rm X} \sim
10^{36-38}$ erg s$^{-1}$) and active galactic nuclei (L$_{\rm X} >
10^{42}$ erg s$^{-1}$), and their extra-nuclear locations, their
physical nature is far from obvious.  Determining this nature is
therefore a compelling challenge.  In this paper we review the
progress made towards this end through recent {\it ROSAT\/}, {\it
ASCA\/} and {\it Chandra\/} studies, and present new {\it Chandra\/}
observations including an X-ray/optical study of two bright ULX, and
an observation of the ULX-rich galaxy pair NGC 4485/90.  We conclude
with a summary of how recent results are advancing our understanding
of the nature of ULX in nearby galaxies.

\section{A new class of black holes?}

A reasonable assumption to make is that ULX are powered by accretion.
If we then consider a ULX emitting isotropically in the X-ray regime
with a luminosity of $\sim 5 \times 10^{39}$ erg s$^{-1}$, a simple
Eddington luminosity limit argument implies that the mass of the
compact accreting object is at least 25 $M_{\sun}$.  If the accretion
is occuring at lower rates, this implies that the compact object mass
is even higher, potentially placing it anywhere in the $10^2 - 10^5
M_{\sun}$ regime (dependent upon the accretion rate).  This is at odds
with previous theoretical and observational studies of black holes
which have found evidence for two main types, the remnants of massive
stars (with $M_{BH} \le 10 M_{\sun}$), and the ``super-massive'' black
holes residing in the nuclei of galaxies ($M_{BH} \ge 10^6 M_{\sun}$).
It is very unlikely that ULX contain the latter super-massive flavour,
as these object would sink to the centre of the host galaxy through
the action of dynamical friction in less than a Hubble time
(c.f. \cite{troberts-E3:tos75}), and the Eddington limit argument
rules out the stellar remnant black holes as insufficiently massive.
It has therefore been suggested that ULX may represent a hitherto
unrecognised $10^2 - 10^5 M_{\sun}$ {\it intermediate-mass\/} class of
black holes (e.g. \cite{troberts-E3:cm99}).

The evidence that a large fraction of ULX may indeed be powered by
accretion onto a black hole is compelling.  Variability studies of ULX
are very informative; both {\it ROSAT\/} and {\it ASCA\/} studies have
detected short-term variations ($\sim 1000$s of seconds) consistent
with accretion processes in the lightcurves of ULX
(e.g. \cite{troberts-E3:zezas99}; \cite{troberts-E3:okada98}).  Other
{\it ASCA\/} studies have observed long-term spectral transitions
between soft/high and hard/ low states, very similar to the state
changes observed in Galactic black hole binaries
(\cite{troberts-E3:lp2001}; \cite{troberts-E3:kubota2001};
\cite{troberts-E3:mizuno2001}). Additionally, the first suggestion of
X-ray periodicity were recently discovered in a deep {\it ASCA\/}
observation of a ULX in IC 342, indicative of it residing in a stellar
binary system ($P \sim 30 - 40$ hours, \cite{troberts-E3:sugiho01}).
{\it ASCA\/} spectroscopy has also provided important evidence towards
a black hole interpretation for ULX, with the spectra of many examples
being well-fit by the ``multi-colour disc black-body'' (MCDBB)
emission model characteristic of an optically thick accretion disc
around a black hole (\cite{troberts-E3:m2000};
\cite{troberts-E3:cm99}).  Intriguingly, this spectrum is also
found to provide a good spectral fit to several Galactic microquasars.
However, the ULX spectral fits are problematic, as the derived
inner-disc temperatures are too high for the massive black holes
expected in these systems. \cite*{troberts-E3:m2000} explain this by
inferring the presence of a rapidly-rotating Kerr black hole, which
allows the inner-edge of the accretion disc to move closer to the
black hole and thus heat up.  An alternate scenario that predicts the
high temperatures observed in ULX, whilst retaining an
intermediate-mass black hole, is the ``slim disc'' accretion regime
(\cite{troberts-E3:wat01}).  In this scenario near Eddington-rate
accretion is occuring and the X-ray emission is dominated by a compact
region located very close to (within three Schwarzschild radii of) the
black hole, which is not required to be rotating.

However, none of the {\it ASCA\/} data make an overwhelming case for
the presence of a $10^2 - 10^5 M_{\sun}$ black hole.  This is in a
large part due to the uncertainty introduced by the source confusion
inherent in the $> 90\arcsec$ spatial resolution of {\it ASCA\/},
which may lead to overestimates of the source luminosities.  The best
evidence for intermediate-mass black holes comes from recent high
spatial resolution {\it Chandra\/} observations.  The starburst
galaxies M82 and NGC 3628 are both observed to host a near-nuclear,
point-like ULX which is seen to be highly variable in observations
spanning periods of several years, and display a peak luminosity of
well over $10^{40}$ erg s$^{-1}$ (e.g. \cite{troberts-E3:kaaret01};
\cite{troberts-E3:stric01}).  This implies black holes with masses in
excess of several hundred $M_{\sun}$ in each ULX.

Recent results have, however, called into question the presence of an
intermediate-mass black hole in many ULX.  {\it Chandra\/}
observations have revealed that large numbers of ULX are present in
very active starburst galaxies with, for example, about ten found in
each of the Antennae and NGC 3256 systems (\cite{troberts-E3:fzm01};
\cite{troberts-E3:lira02}).  Other ULX, such as those in M82 and NGC
3628 mentioned above, also appear in star forming regions.  {\bf This
implies that a large proportion of the ULX population is intrinsically
linked to active star formation.\/} This may be inconsistent with the
presence of an intermediate-mass black hole, with the strongest
objection being that most formation scenarios, such as their creation
from the hierarchal merging of black holes in the centre of a globular
cluster, require considerably longer than the $\sim 10^8$ year
lifetime of a starburst event. Also, this model would locate the
sources very close to the nucleus of the galaxy where the potential
well is deep enough to retain the stellar mass black holes within the
cluster (see \cite{troberts-E3:king01} and references therein for
further discussion).  We must therefore consider alternative origins
for the ULX phenomenon.

One possibility is that a reasonable proportion of the ULX population
is composed of recent supernovae, which are known to reach X-ray
luminosities of up to $10^{41}$ erg s$^{-1}$ if they explode in dense
environments (e.g. SN 1988Z, \cite{troberts-E3:ft96}).  Of the 28 ULX
catalogued by \cite*{troberts-E3:rw2000}, 3 -- 4 are indeed
identifiable with recent supernovae.  However, given the spectral and
variability characteristics of ULX described above it is unlikely that
a much larger fraction have this origin.  Instead, it may be that a
large proportion of ULX are mildly-beamed X-ray binaries, as suggested
by \cite*{troberts-E3:king01}.  Crucially, if ULX emit anisotropically
then this removes the necessity for an intermediate-mass black hole.
It does however imply large numbers of ULX, as only a small fraction
of a beamed population will direct their emission along our line of
sight at any one time.  This problem is solved if the beaming
originates in ``ordinary'' black hole and/or neutron star
intermediate- and high-mass X-ray binaries.  This scenario produces a
viable solution to the problem of associating ULX with ongoing star
formation, as the short lifetime of high-mass X-ray binaries is well
matched to the lifetime of the starburst.  Hence this is a promising
candidate for explaining the nature of a large proportion of ULX.

\cite*{troberts-E3:king01} also postulate that the ULX phase is
associated with a short-lived but common epoch of thermal-timescale
mass transfer, which is inevitable in intermediate- and high-mass
X-ray binaries.  This in turn provides a potential physical link to
the Galactic microquasars.  The possibility of this physical
similarity is further discussed by \cite*{troberts-E3:geor02}, who
conclude that ULX are likely to be microquasars located in nearby
galaxies that are observed with their jet oriented into our
line-of-sight (a phenomenon that may also referred to as
``microblazars''; see
\cite{troberts-E3:mr99}).  Hence we now have a viable alternative to
the intermediate-mass black hole scenario which may be tested
observationally by looking for further similarities between the ULX
and Galactic microquasar phenomena.

\section{An X-ray/optical study of two ULX}

The ULX phenomenon is comparatively poorly studied, mainly as result
of their moderate observable X-ray fluxes ($\sim 10^{-13} - 10^{-12}$
erg cm$^{-2}$ s$^{-1}$ for the limited number of ULX within 10 Mpc)
and the small X-ray collecting areas of previous missions.  ULX are
also particularly poorly understood in a multi-wavelength context,
with little or no evidence for discrete counterparts previously
reported in the literature.  However, the increased effective areas
and better spectral and spatial resolution of {\it Chandra\/} and {\it
XMM-Newton\/} now provide us with a first opportunity to study many of
these objects in detail.

For our study we adopted a two-pronged approach, obtaining {\it
Chandra\/} X-ray data and William Herschel Telescope/INTEGRAL optical
data.  The {\it Chandra\/} ACIS-S data\footnote{Note that the ACIS-S
data for the two observations discussed below were obtained in a
sub-array mode to mitigate the effects of detector pile-up.} were
obtained to provide the best indication of whether the ULX are truly
point-like at the highest currently available X-ray spatial
resolution, and to provide the subarcsecond astrometry critical to
follow-up the X-ray sources with multi-wavelength studies.  The data
were obtained in two epochs, separated by several months, to provide
an initial study of the X-ray characteristics of the ULX, and their
gross variability.  Optical data were obtained with the ``INTEGRAL''
integral field unit on the William Herschel Telescope, La Palma.  This
provided 189 separate optical spectra covering a $16\arcsec \times
12\arcsec$ field of view, over which images can be reconstructed
(utilising the relative fibre positions) in any narrow-band within the
$\sim 4500 - 7500$ \AA~ wavelength coverage of the observations.
Here, we detail the results of this study for two of the nearest and
brightest ULX catalogued by \cite*{troberts-E3:rw2000}.

\subsection{NGC 4559 X-1}

\begin{figure}
\centering
\includegraphics[width=8cm]{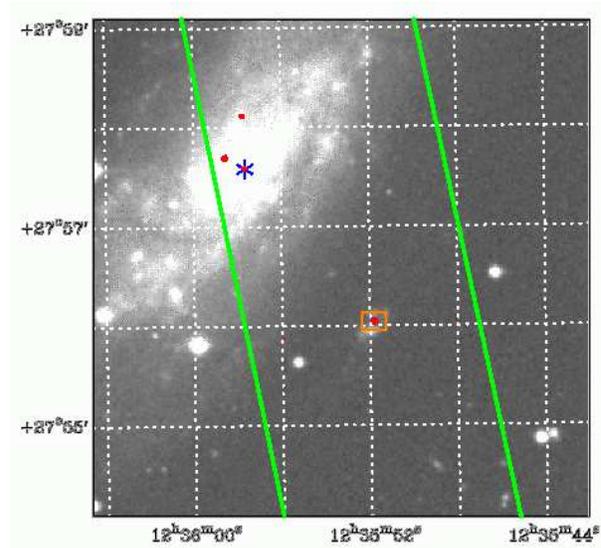}
\caption{The position of NGC 4559 X-1 relative to its host galaxy.
The optical greyscale image is DSS-2 blue data.  Overlaid onto the
image are {\it Chandra\/} ACIS-S X-ray contours (in red), the coverage
of the {\it Chandra\/} ACIS-S sub-array (green) and the field-of-view
of the INTEGRAL instrument (orange).  The nucleus of NGC 4559 is
marked by an asterisk, and hosts a discrete X-ray source.} 
\end{figure}

\begin{figure}
\centering
\includegraphics[width=6cm,angle=270]{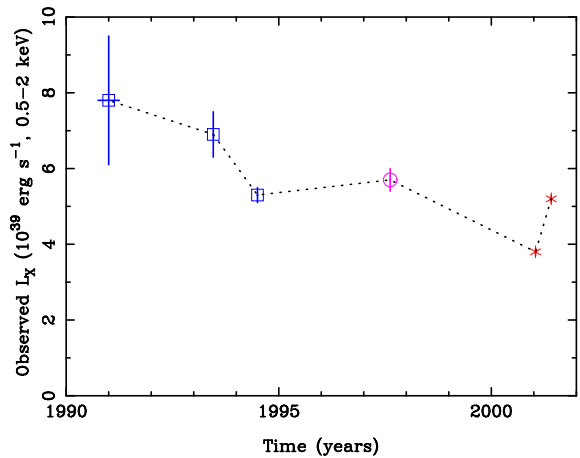}
\caption{Long-term X-ray variability of NGC 4559 X-1.  {\it ROSAT\/}
PSPC data points are shown in blue; {\it ROSAT\/} HRI in magenta; and
{\it Chandra\/} ACIS-S in red.  The lightcurve is derived in the 0.5 -
2 keV band as this range is common to all three instruments.}
\end{figure}

This ULX was first identified by \cite*{troberts-E3:vpb97}, who
inferred from {\it ROSAT\/} PSPC data that it was likely to be a
several hundred year old buried supernova remnant on the outskirts of
NGC 4559 (note that they catalogue it as NGC 4559 X-7, but we use the
nomenclature of \cite{troberts-E3:rw2000}).  Its position is shown in
Figure 1, where it appears coincident with an anomalous group of
H{\small II} regions on the outskirts of the galaxy (see also
\cite{troberts-E3:pm02}, and below).  Its X-ray profile is entirely
consistent with a point-like X-ray source, and it is very luminous;
the {\it Chandra\/} ACIS-S data measure a luminosity of $\sim 10^{40}$
erg s$^{-1}$ in both observation epochs.  There is no significant
short-term X-ray variability observable in either epoch, though a
long-term lightcurve based on {\it ROSAT\/} data and the current
observations suggests a gradual (though not linear) reduction in the
flux level over the last 10 years (see Figure 2).  The analysis of the
X-ray spectra of the ULX proved problematic, with no simple models
providing very good fits.  However, the best fits were achieved with a
simple powerlaw model, which showed the ULX spectrum to soften
slightly as the flux increased between the two observation epochs
(from $\Gamma \approx 1.9$ to 2.15, with no corresponding change in
absorption column, over six months).  This behaviour is similar to
that observed previously in other ULX, where the spectra appears
softer in higher flux states, and is consistent with the
interpretation of NGC 4559 X-1 as a black hole X-ray binary system.

\begin{figure*}
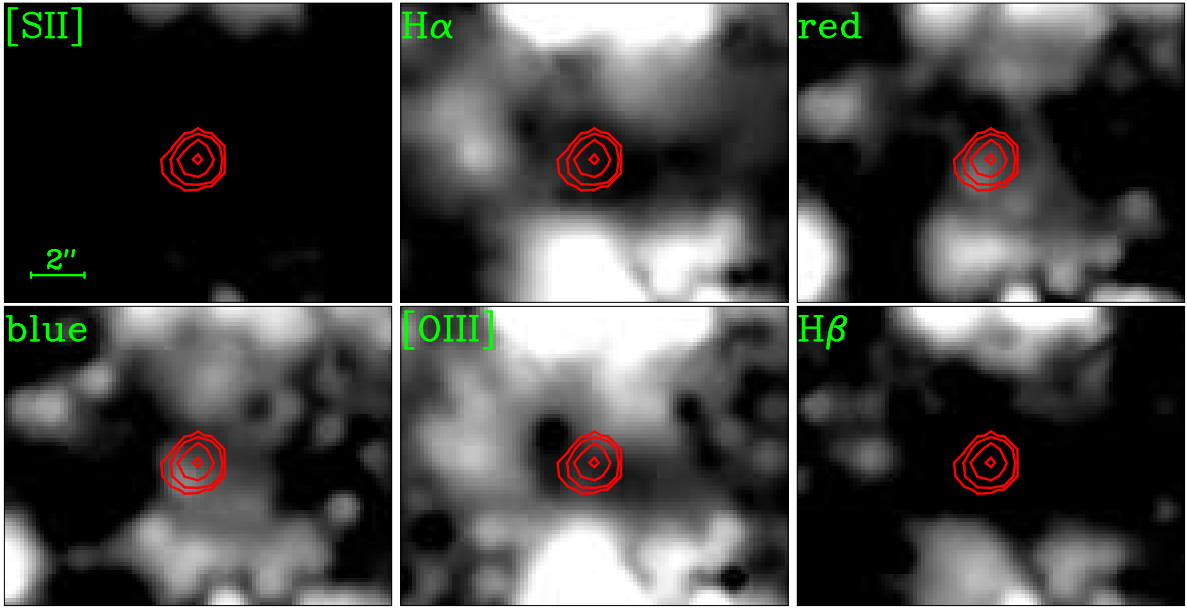

\centering
\includegraphics[width=4cm,angle=270]{troberts-E3_fig3a.ps}
\includegraphics[width=4cm,angle=270]{troberts-E3_fig3b.ps}
\includegraphics[width=4cm,angle=270]{troberts-E3_fig3c.ps}
\vspace*{0.4cm}
\includegraphics[width=4cm,angle=270]{troberts-E3_fig3d.ps}
\includegraphics[width=4cm,angle=270]{troberts-E3_fig3e.ps}
\includegraphics[width=4cm,angle=270]{troberts-E3_fig3f.ps}
\caption{INTEGRAL narrow-band images of the region containing NGC 4559 
X-1.  The bands are labelled in green, and ordered in decreasing
wavelength from the top left.  The emission-line bands are
continuum-subtracted, and all bands have the same intensity scaling
for direct comparison.  Arbitrary {\it Chandra\/} ACIS-S contours
shown the position of the ULX (in red).}
\end{figure*}

In Figure 3 we show the INTEGRAL data for the region containing NGC
4559 X-1.  This is shown as a series of four continuum-subtracted
emission-line images, in the [S{\small II}], H$\alpha$, [O{\small
III}] and H$\beta$ bands respectively, and also red continuum (6000 --
6200 \AA) and blue continuum (5100 -- 5200 \AA) bands for comparison.
The data shows that there is no obvious optical counterpart to NGC
4559 X-1.  Instead, the most interesting feature of the images is the
location of NGC 4559 X-1 in the centre of an apparent cavity in a
region of emission-line gas surrounding the ULX.  This leads to an
interesting question: is NGC 4559 X-1 responsible for the
emission-line nebulae?  The nebulae have previously been identified as
a series of H{\small II} regions (\cite{troberts-E3:vpb97}), and hence
assumed to be powered by young stars, which would of course fit
conveniently into the emerging relationship of ULX with star
formation.  However, the position of the ULX in the centre of the
nebulae may call this into question, as it suggests the ULX is somehow
responsible for the nebulae.  A study of the extent of the influence
of NGC 4559 X-1 on the surrounding nebulae is beyond the scope of this
paper, but it is interesting to note that while this system may
possibly be another manifestation of the relationship between ULX and
active star formation, this cannot be established for certain until we
distinguish to what degree the nebulae are energised by the ULX or the
putative young stars.  A further point is that it is not entirely
clear from the data whether the morphology of the surrounding gas is
ring-like or conical.  If it is conical, and a result of irradiation
by the ULX, then this clearly supports the model of ULX as beamed
systems.

\subsection{NGC 5204 X-1}

This ULX was first detected in {\it EINSTEIN\/} data
(\cite{troberts-E3:fkt92}), and has since been catalogued in several
{\it ROSAT\/} surveys, but has never been well-studied to date.  The
{\it Chandra\/} ACIS-S data reveal NGC 5204 X-1 to lie $\sim
15\arcsec$ away from the nucleus of its host galaxy, the
Magellanic-type galaxy NGC 5204 (see Figure 4).  It is also point-like
at the resolution of {\it Chandra\/}, and has a luminosity that varies
in the range $2 - 5 \times 10^{39}$ erg s$^{-1}$ between the two
observations (an interval of four months).  Again, we see no evidence
of short-term variability in the data, but the long-term lightcurve
shows evidence for strong variability over a baseline of 20 years
(Figure 5).  The X-ray spectrum is again best-fit by a powerlaw
continuum model in both epochs.  The best fits show that as the flux
drops to $\sim 40\%$ of its original value over the four months
interval between observations, the spectral index of the powerlaw
softens from $\Gamma \approx 2.4$ to $\Gamma \approx 2.9$.  This is
counter to the behaviour shown by other ULX and black hole X-ray
binaries in general.  However, this unusual behaviour may constitute
the best evidence yet linking ULX to microquasars, as an anomalous
soft/low state is also observed by {\it XMM-Newton\/} for the Galactic
microquasar GRS 1758-258 (\cite{troberts-E3:mil02}).

\begin{figure}
\centering
\includegraphics[width=8cm]{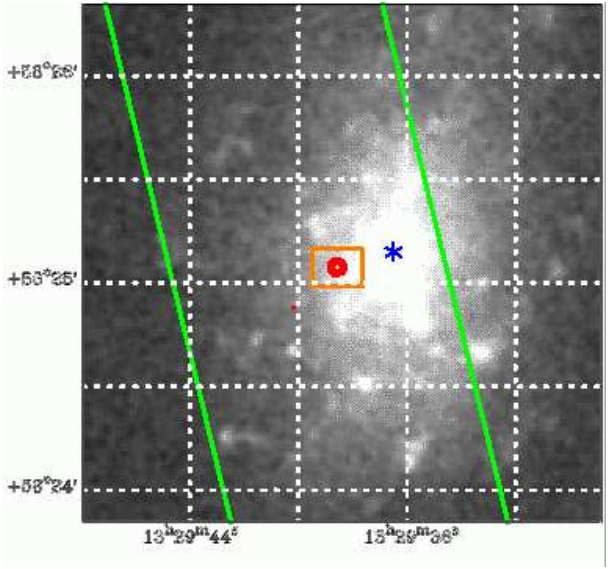}
\caption{The position of NGC 5204 X-1 relative to its host galaxy.
The figure is arranged similarly to Figure 1.}
\end{figure}

\begin{figure}
\centering
\includegraphics[width=6cm,angle=270]{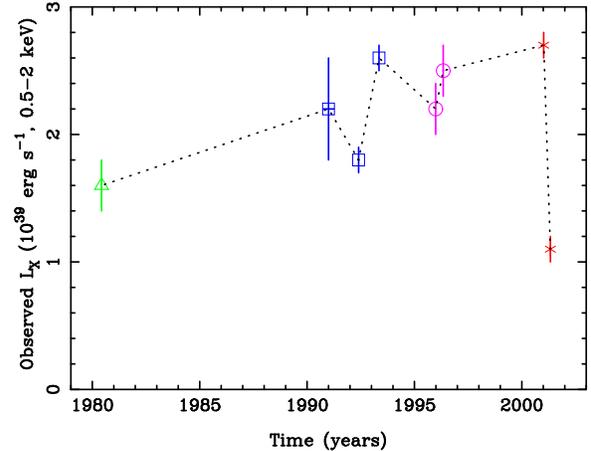}
\caption{Long-term X-ray variability of NGC 5204 X-1, displayed similarly
to Figure 2, but with the addition of an {\it EINSTEIN\/} IPC data
point (shown in green).}
\end{figure}

\begin{figure*}
\centering
\includegraphics[width=4cm,angle=270]{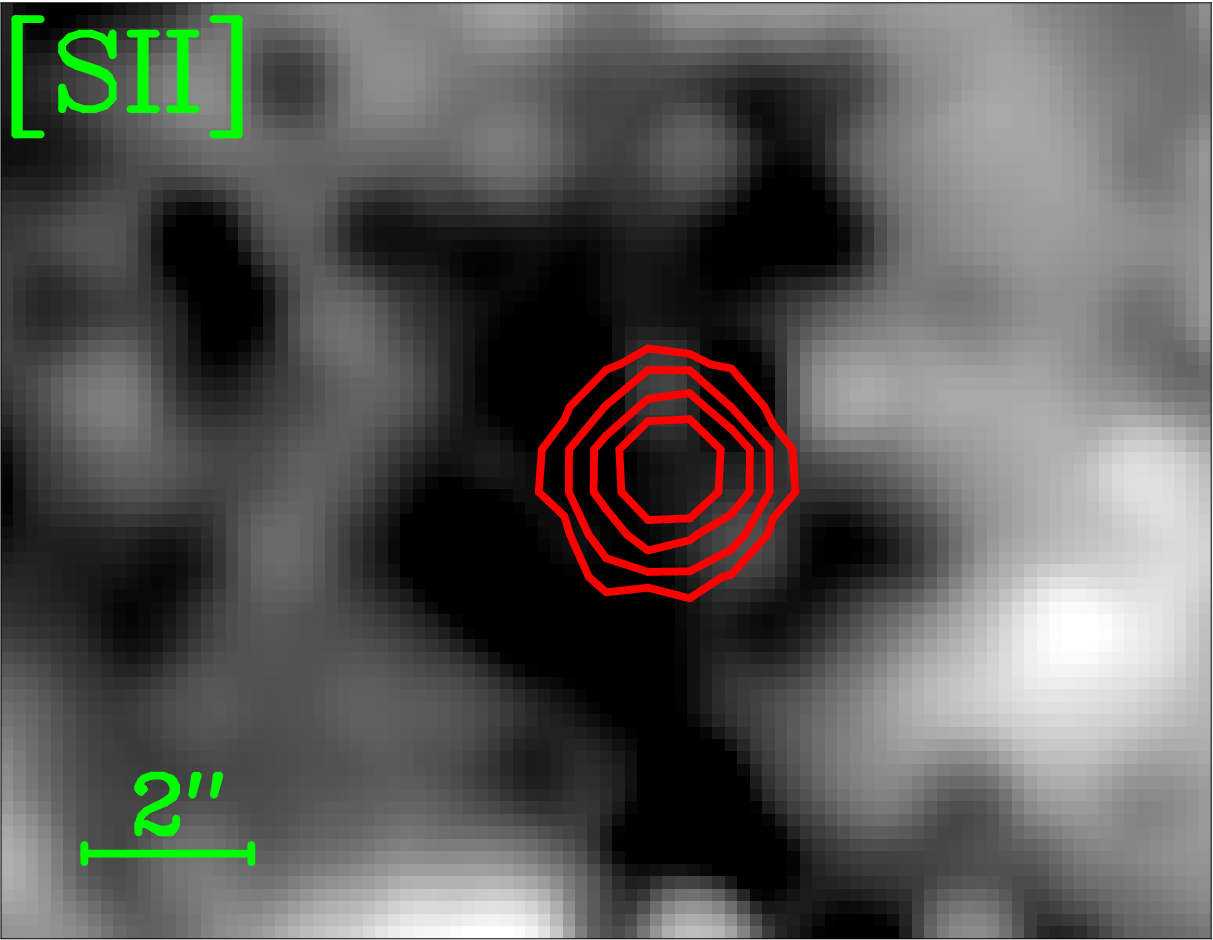}
\includegraphics[width=4cm,angle=270]{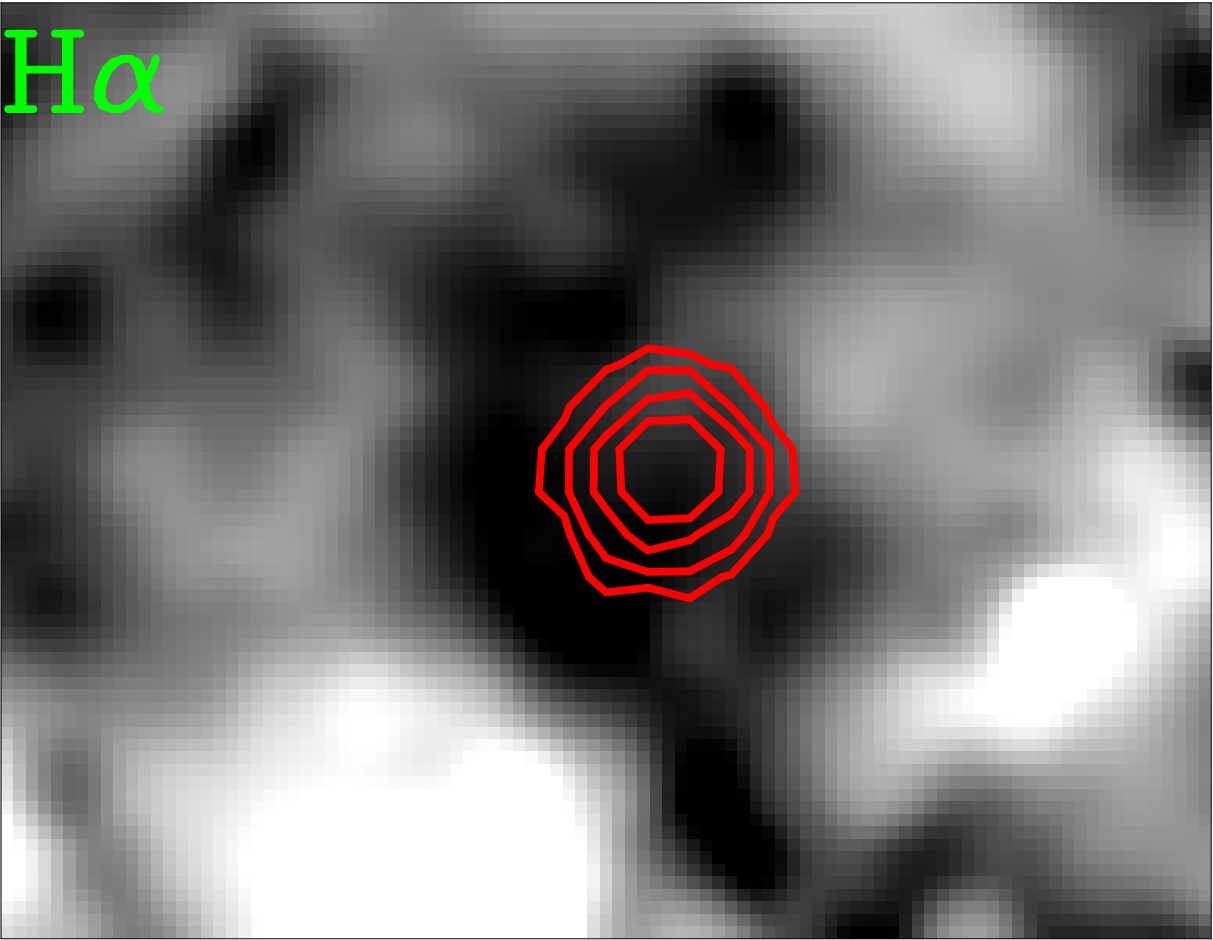}
\includegraphics[width=4cm,angle=270]{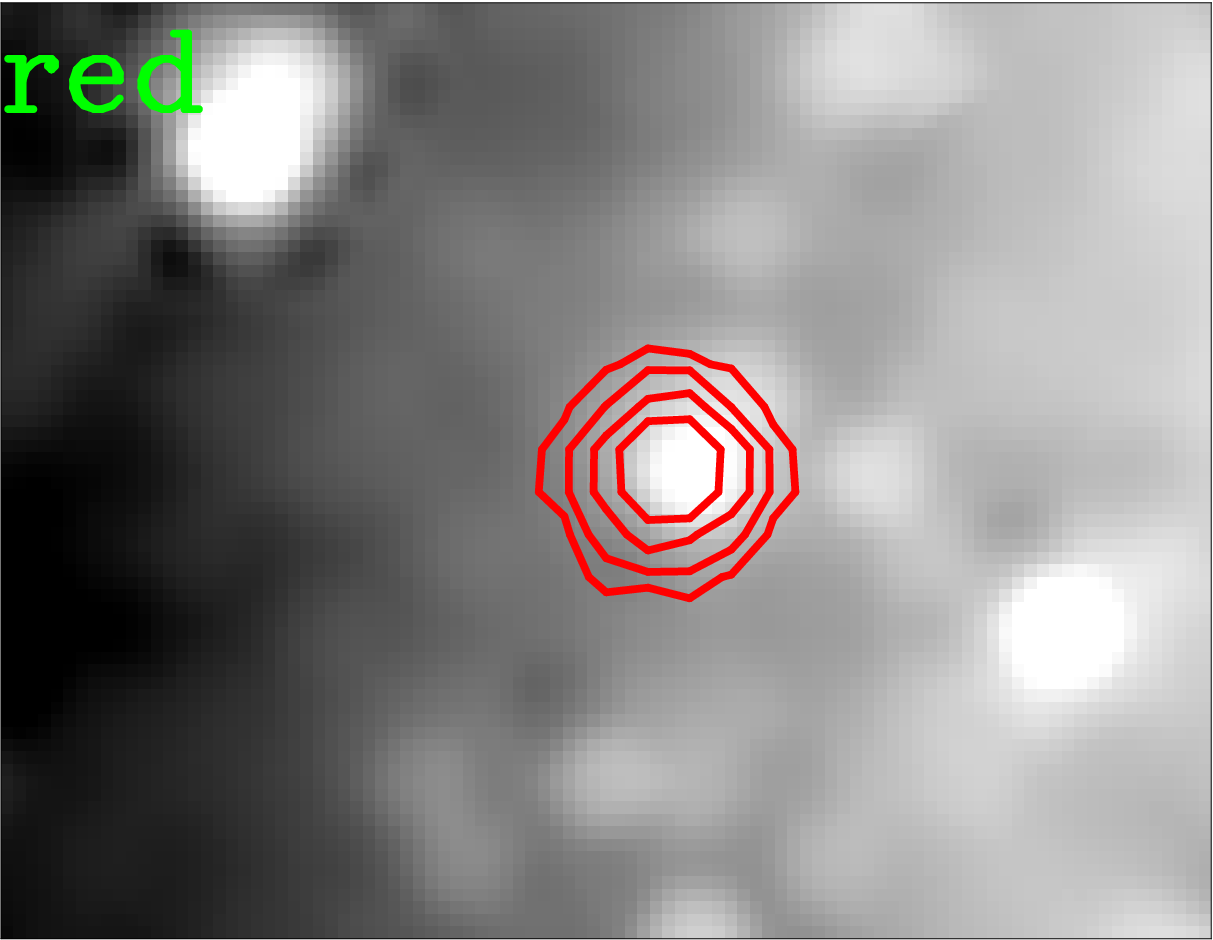}
\vspace*{0.4cm}
\includegraphics[width=4cm,angle=270]{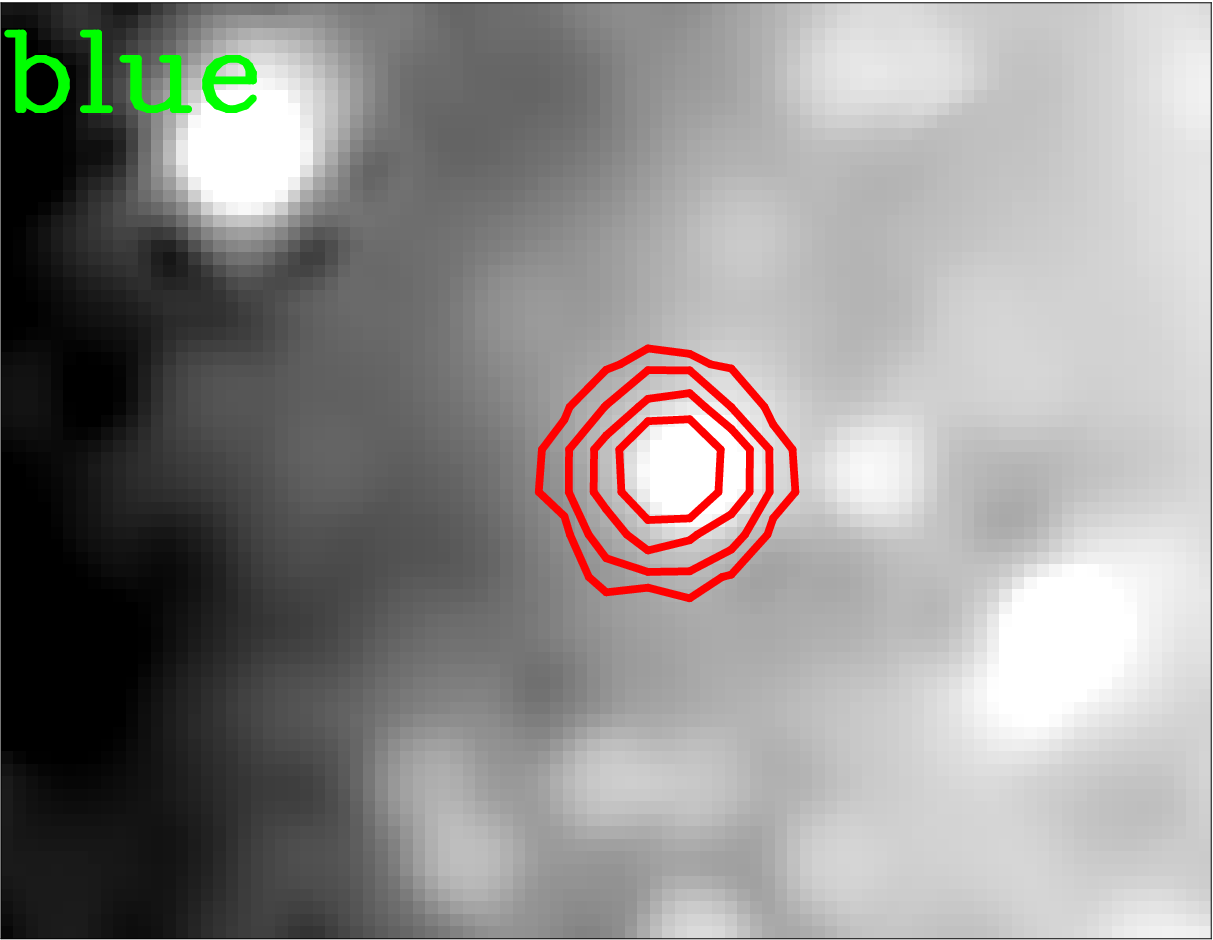}
\includegraphics[width=4cm,angle=270]{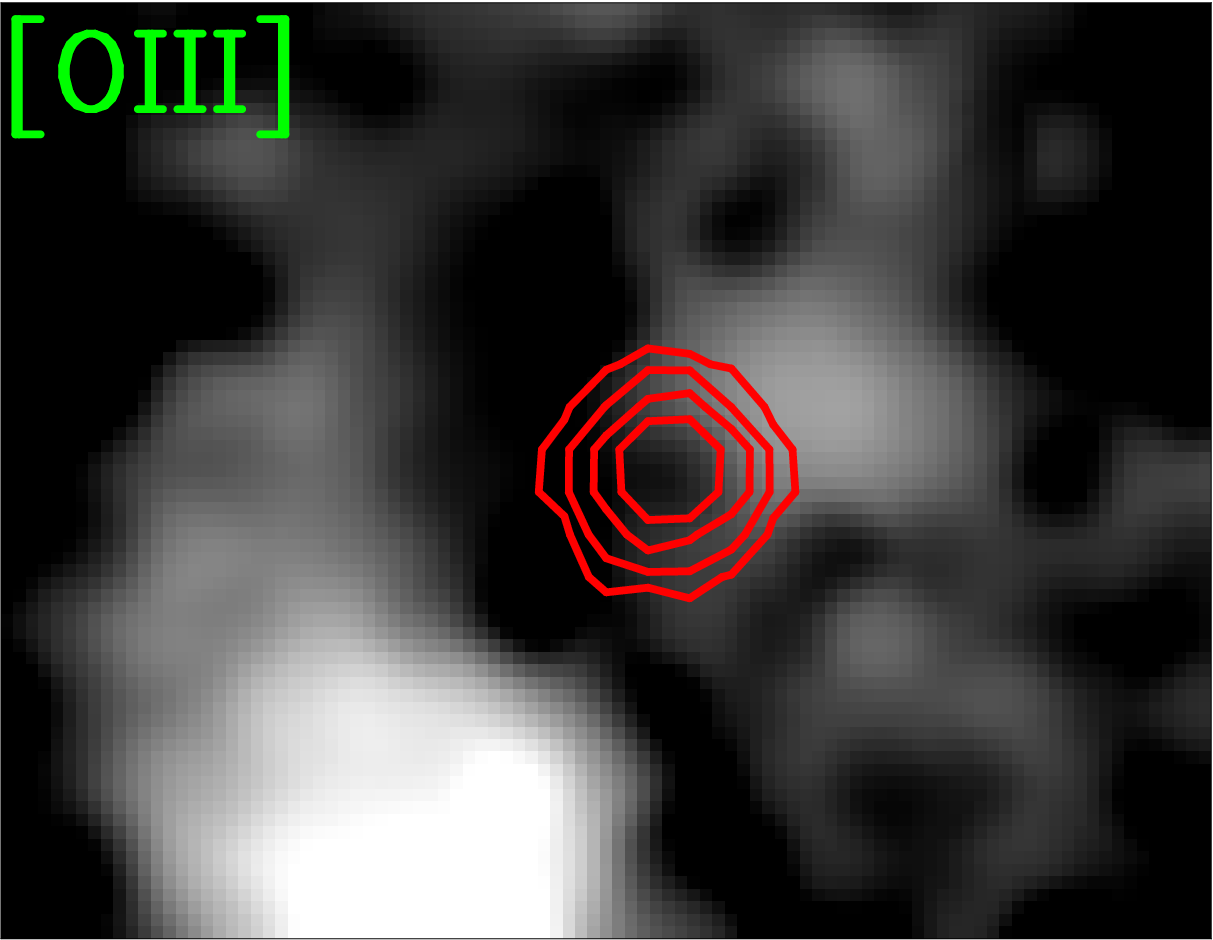}
\includegraphics[width=4cm,angle=270]{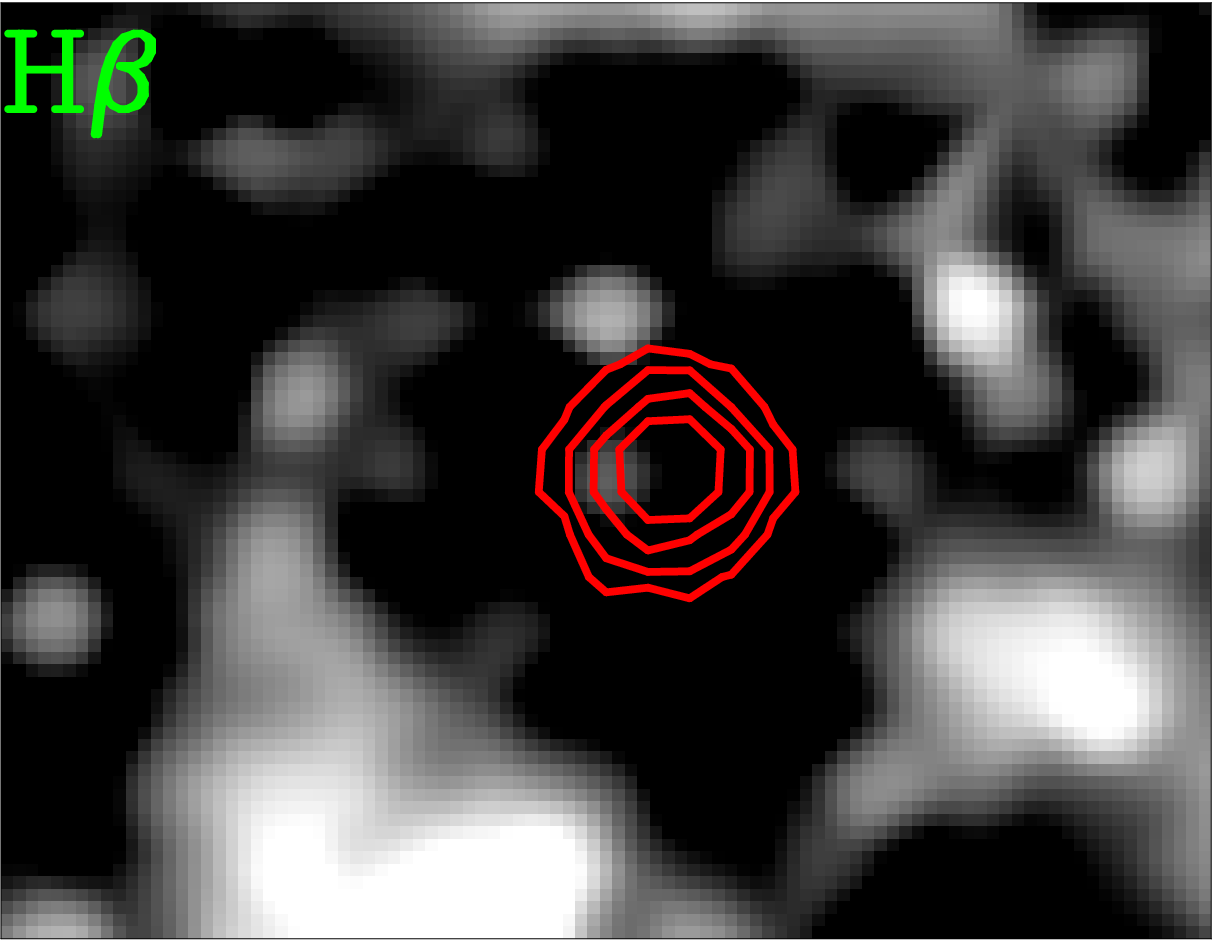}
\caption{INTEGRAL narrow-band images of the region containing NGC 5204 
X-1, displayed similarly to Figure 3, albeit with the image intensities
arbitrarily-scaled.}
\end{figure*}

Perhaps the most important discovery in this work was reported in
\cite*{troberts-E3:rob01}, in which we revealed the detection of the
first possible stellar optical counterpart to an ULX.  This object is
dominated by blue continuum emission, and so only appears in the
continuum band images shown here in Figure 6, and is discussed below.
NGC 5204 X-1 also appears to be located in the centre of a cavity
evident in the continuum-subtracted emission-line images (though we
note that this is not as ``clean'' an environment as NGC 4559 X-1,
since NGC 5204 X-1 is located further within its galaxy where source
confusion is a greater issue).  This potential similarity to the
environment of NGC 4559 X-1 raises the question of whether a large
number of ULX are located in similar cavities, and implies that they
may be a common feature of ULX.  Further results on this phenomenon,
showing it may indeed be common to ULX, are presented by
\cite*{troberts-E3:pm02} in these proceedings.

In \cite*{troberts-E3:rob01} we discussed the nature of the optical
counterpart, concluding from its featureless blue continuum optical
spectrum that it was likely to be an O-star, or a small O-star
association, in NGC 5204 (since at $m_v = 19.7$ we require at least 5
O supergiants to be present to account for the optical flux).
However, we could not completely rule out the scenario in which the
X-ray source and its optical counterpart are actually a BL Lac object
located behind the galaxy.  This can now be ruled out.  Firstly, newly
public {\it HST\/} WFPC images covering the region around NGC 5204 X-1
resolve the previous counterpart into two separate, yet still
point-like, optical sources at $0.1\arcsec$ resolution, both of which
remain blue in colour.  The resolved sources are shown in Figure 7.
The brighter of the sources has $m_v \sim 20.5$, and is still
consistent with at least four O supergiants in NGC 5204.  Secondly,
new VLA radio data has placed a stringent limit of 84 $\mu$Jy on the
3.6 cm radio continuum emission at the position of NGC 5204 X-1
(\cite{troberts-E3:wong02}, these proceedings), which implies the
radio flux is too low for the source to be identified as a BL Lac.  We
are therefore now confident that we are observing a good candidate ULX
and optical counterpart system in NGC 5204.  The O-star classification
is wholly consistent with the relationship between ULX and star
formation, and indeed is consistent with the
\cite*{troberts-E3:king01} scenario in which ULX are ordinary
high-mass X-ray binaries with X-ray emission beamed into our
line-of-sight.  Furthermore, the noted similarity between the soft/low
state of NGC 5204 X-1 and that observed in the Galactic microquasar
GRS 1758-258 may provide the some of the best evidence so far that ULX
and microquasars are one and the same phenomenon.

\section{The ULX population of NGC 4485/4490}

A further, perhaps more convenient, method of studying the X-ray
properties of ULX is to observe several of them at once.  This is only
possible in a few galaxies in the local universe, which tend to be
systems undergoing very active star-formation (see Section 2).  One
such host is the tidally-interacting late-type galaxy pair NGC
4485/NGC 4490 which is located at a distance of only 7.8 Mpc. It was
observed to host three ULX in the {\it ROSAT\/} HRI survey of
\cite*{troberts-E3:rw2000}.  A 20 ks {\it Chandra\/} ACIS-S GTO was
undertaken in November 2000, the results of which are discussed in
\cite*{troberts-E3:rob02}.  Here, we concentrate on the remarkable
population of ULX unveiled by this observation.

\begin{figure}
\centering
\includegraphics[width=8cm]{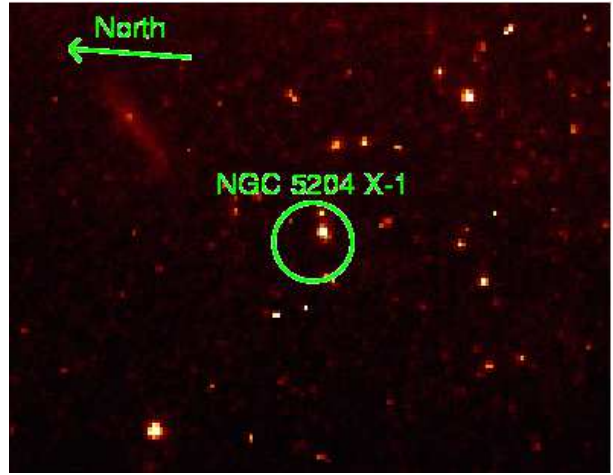}
\caption{{\it HST\/} WFPC F606W filter image covering the region
around NGC 5204 X-1.  North is indicated.  The error circle shown
around the {\it Chandra\/} ACIS-S position is 1\arcsec in radius.}
\end{figure}

\begin{figure}
\centering
\includegraphics[width=9cm]{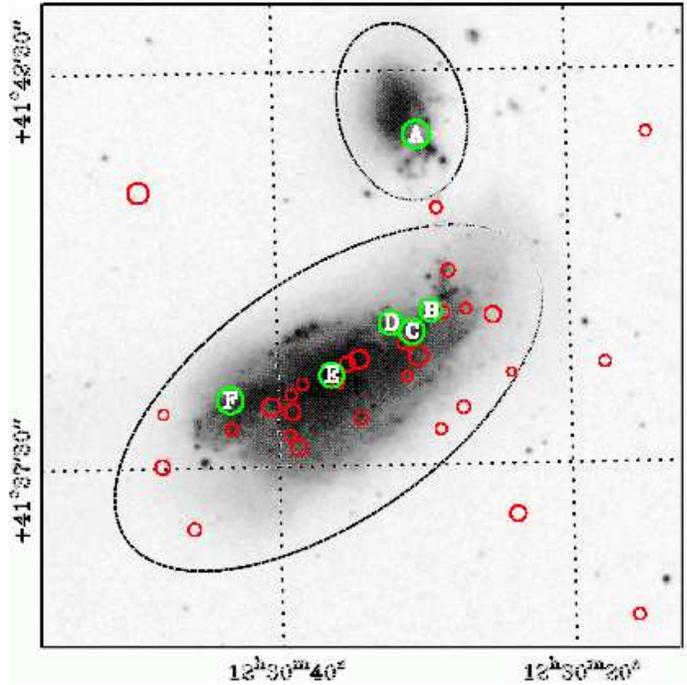}
\caption{{\it Chandra\/} ACIS-S X-ray source detections in the NGC
4485/90 field overlaid onto a greyscale DSS-2 blue image of the
galaxies.  NGC 4490 is the larger, southern galaxy and contains the
majority of the X-ray source detections.  The six ULX are indicated by
green circles and labelled A - F.  A further 24 X-ray sources are
found within the optical extent of the galaxies (i.e. within each
D$_{25}$ ellipse), with luminosities down to 3 $\times$ 10$^{37}$ erg
s$^{-1}$, and are shown by the red circles.  The circle sizes scale
logarithmically with flux for each source.}
\end{figure}

\begin{figure*}
\centering
\includegraphics[width=12cm,angle=270]{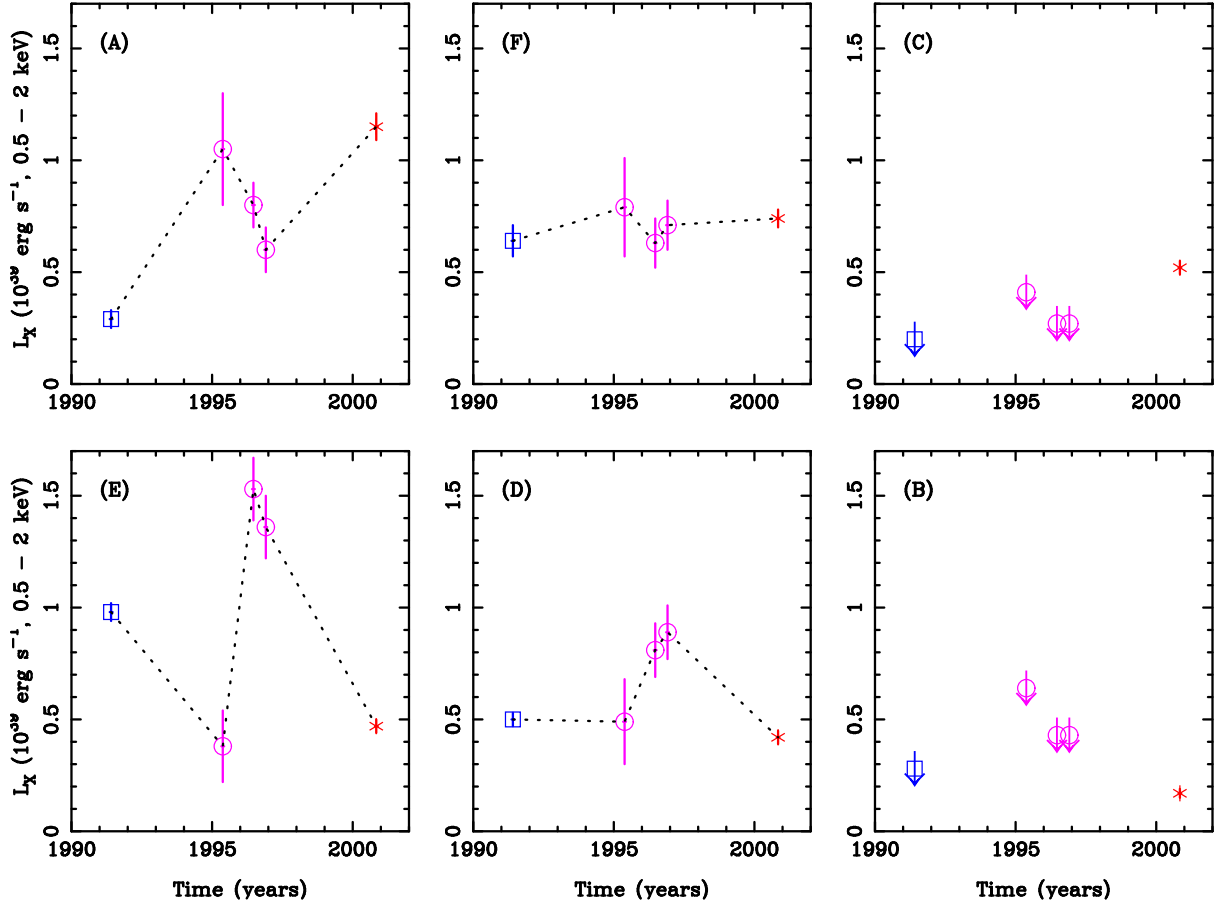}
\caption{Long-term lightcurves of the six ULX in NGC 4485/90.
The formatting is as per Figures 2 \& 5, except for the upper limits
shown by downwards-pointing arrows.  The labels (A) - (F) refer to the
sources as labelled in Figure 8.}
\end{figure*}

The {\it Chandra\/} ACIS-S observation reveals a total of six ULX
within NGC 4485/90, ranging in observed luminosity between $\sim 1$
and $4 \times 10^{39}$ erg s$^{-1}$ (0.5 - 8 keV).  The positions of
the ULX within the galaxies are shown in Figure 8.  Note that we see
all three ULX reported in \cite*{troberts-E3:rw2000}, and of the three
new ULX two appear due to the hard ($> 2$ keV) response of the ACIS-S
(one of which was previously unknown, the other known but not luminous
enough to be classified as an ULX).  The third new ULX appears to be a
truly transient source.  An examination of the short-term variability
data for each ULX shows that none are significantly variable over the
duration of the {\it Chandra\/} observation.  However, the long-term
lightcurves (Figure 9) show that significant variability characterises
four of the six ULX over the 10-year observation baseline.  Three of
the four sources detected at all five observation epochs are strongly
variable (A, D and E); only source F shows little or no variability.
Source C is the transient, with {\it ROSAT} (0.5 - 2 keV) flux upper
limits below the observed {\it Chandra\/} flux.  Only one ULX is too
faint to have appeared in any previous observation, source B.  An
inspection of its spectral data shows this is because it is absorbed
by a column of well in excess of $10^{22}$ atom cm$^{-2}$.  However,
it actually has the softest intrinsic spectral form of the six ULX (a
thermal bremsstrahlung with $kT \sim 1.5$ keV).  It is also exactly
coincident with a FIRST survey radio source, suggesting its
identification as a recent supernova in NGC 4490.

The other five ULX all have intrinsically harder spectra, though are
subject to less absorption column (which varies between 2 and $9
\times 10^{21}$ atom cm$^{-2}$ over the five sources).  The two brightest
ULX (A and F) are best fit by the MCDBB model favoured in many {\it
ASCA\/} spectral fits of ULX.  The other three ULX, for which the
spectral data are of a lower quality, are all equally well-fit by
either an MCDBB or a powerlaw continuum model.  However, the best-fit
values given by the MCDBB model all have values of $kT_{in}$ in the
$1.2 - 1.8$ keV range found in {\it ASCA\/} spectroscopy.  It appears,
therefore, that on the basis of their spectral and temporal
characteristics the five ULX without radio counterparts are all
consistent with black-hole X-ray binaries.

The presence of six ULX in a small galaxy pair is quite remarkable, as
the non-nuclear X-ray source luminosity distribution of
\cite*{troberts-E3:rw2000} predicts that we should only observe $\sim
0.5$ ULX in the NGC 4485/90 system, based on the combined $L_B$ of NGC
4485/90.  However, this system shows considerable evidence for
enhanced star formation in both radio and far infra-red observations,
with a star formation rate of $\sim 5 M_{\sun}$ per year
(\cite{troberts-E3:cag99}).  The ongoing star formation is mainly
located in the tidal arms and bridge between the galaxies, with
further peaks of activity slightly to the west of the nucleus of NGC
4490 and in a northern spiral arm.  The six ULX are all present within
these regions, with source A located in the tidal tail of NGC 4485,
sources B, C and D in the western (tidally disrupted) spiral arm of
NGC 4490, source E in the near-nuclear star forming region and source
F in the northern spiral arm (c.f. Figure 8).  {\bf This direct
spatial co-location of the ULX and the star formation regions in NGC
4485/90 is more confirmation of the apparently strict relationship
between the phenomena.}  Additionally, the inferrence that five of the
six ULX are black hole X-ray binaries implies that even in very active
star formation regions, the ULX phenomenon is predominantly due to
accreting sources as opposed to supernovae.

\section{Conclusions}

Our knowledge of the properties of ULX is growing rapidly in the era
of {\it Chandra\/} and {\it XMM-Newton\/}.  The following themes are
emerging:

\begin{itemize}
\item
ULX are a heterogeneous class, displaying a variety of different
temporal and spectral X-ray characteristics, though with broad
similarities in many cases.
\item
A small number (10 - 20\% of ULX?) are identifiable with recent
supernovae, and the remainder appear to have the X-ray characteristics
of black hole X-ray binaries.
\item
A pattern of the spatial co-location of ULX with active star formation
regions is emerging, demonstrated clearly in the case of NGC 4485/90,
implying a direct link between the presence of ULX and the ongoing
star formation.
\item
A viable model for the nature of ULX that links the two preceding
points is that ULX may be high-mass X-ray binaries with their X-ray
emission beamed into our line-of-sight.  The case of NGC 5204 X-1
presented above strongly supports this scenario, in that we have an
optical counterpart that appears to be one or more O-stars, along with
possible evidence that the ULX has a rare X-ray spectral state that is
similar to that observed in a Galactic microquasar (which could
provide a physical basis for the beaming hypothesis).
\item
A new phenomenon highlighted by our work is the presence of apparent
cavities in emission-line gas immediately in the vicinity of the ULX
NGC 5204 X-1 and NGC 4559 X-1.  This immediately raises questions as
to the rate of occurrence and the formation and energetics of these
nebulae, which will be the subject of future work.
\item
Though much circumstantial evidence now points to a model of ULX as
beamed high-mass X-ray binaries, it is important to remember that
there is as of yet no clear and unambiguous evidence to prove that
they are not intermediate-mass black holes.
\end{itemize}

It is notable that in the case of NGC 5204 X-1 the key piece of
evidence in linking it to Galactic microquasars is its spectral
variability.  Since the majority of the ULX with established long-term
lightcurves appear highly variable, it may be that their long-term
behaviour between separate epochs will provide the key diagnostics for
establishing their true natures.  This will be tested with future
observations (and re-observations!) of the nearest ULX with both {\it
Chandra\/} and {\it XMM-Newton\/}.  The EPIC instruments on {\it
XMM-Newton\/} offer the best opportunity to constrain the X-ray
spectral characteristics of ULX, and the simultaneous UV imaging with
the Optical Monitor will allow an unprecedented view of the link
between ULX and the young, hot stars associated with stellar formation
regions.

\begin{acknowledgements}

TPR, MRG \& PL gratefully acknowledge funding from PPARC.  TPR also
thanks Andrew King for very useful discussions.  This paper uses DSS-2
data extracted from the ESO online digital sky survey.  The archival
{\it ROSAT\/} data were obtained from the Leicester database and
archive service (LEDAS) at the University of Leicester.

\end{acknowledgements}

\end{document}